# Robust Coding for Lossy Computing with Receiver-Side Observation Costs

Behzad Ahmadi and Osvaldo Simeone
Department of Electrical and Computer Engineering
New Jersey Institute of Technology
University Heights, Newark, New Jersey 07102
Email: ba63@njit.edu, osvaldo.simeone@njit.edu

*Abstract*—[1]An encoder wishes to minimize the bit rate necessary to guarantee that a decoder is able to calculate a symbol-wise function of a sequence available only at the encoder and a sequence that can be measured only at the decoder. This classical problem, first studied by Yamamoto, is addressed here by including two new aspects: (*i*) The decoder obtains noisy measurements of its sequence, where the quality of such measurements can be controlled via a cost-constrained "action" sequence; (*ii*) Measurement at the decoder may fail in a way that is unpredictable to the encoder, thus requiring robust encoding. The considered scenario generalizes known settings such as the Heegard-Berger-Kaspi and the "source coding with a vending machine" problems. The rate-distortion-cost function is derived and numerical examples are also worked out to obtain further insight into the optimal system design.

## I. Introduction

A common problem in applications ranging from sensor networks to cloud computing is that of calculating a function of the data available at distributed points of a communication network. The aim is typically that of finding the most effective way to operate the network in terms of the resources need for exchanging or collecting information. This general problem is studied in the context of different disciplines, most notably computer science [1] and information theory [2, Chapter 22]. While the computer science literature typically focuses on the calculation of a single instance of the given function, information theory concentrates on the repeated calculation of the function over data sequences.

Adopting the information-theoretic viewpoint, the baseline problem of interest is illustrated in Fig. 1. Here, an encoder measures a sequences $X^n = (X_1, ..., X_n)$, while a decoder measures a correlated sequence $Y^n$. The goal is to minimize the number of bits that the encoder needs to send to the decoder so as not enable the latter to compute a function $T_i = f(X_i, Y_i)$ for all $i = 1, ..., n$, denoted as $T^n = f^n(X^n, Y^n)$, within a given distortion. This problem, which we refer to as *lossy computing*, was first studied in [3], where a general solution was provided, along with specific examples for binary sources and functions. Imposing a zero-distortion constraint, the problem was further studied in [4], where a more compact solution was obtained exploiting a graph-theoretic formulation

[1]This work has been supported by the U.S. National Science Foundation under grant CCF-0914899.

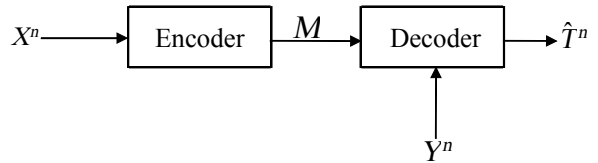

Fig. 1. Lossy compuitng where decoder wishes to compute a function $T^n = f^n(X^n, Y^n)$ with $T_i = f(X_i, Y_i)$ for $i \in [1, n]$.

(see also [2, Chapter 22] for a review). Extensions to a network scenario are studied in [5].

In this paper, we address the lossy computing problem by including two novel aspects that are motivated by the applications to sensor networks and cloud computing. Specifically, we introduce:

1) *Observation costs*: In sensor networks, acquiring information about the data sequence typically consumes energy resources. When such resources are at a premium, it becomes imperative to perform measurements in the most cost-effective way, while still meeting the application requirements. Incidentally, this is the same principle that motivates compressive sensing. As illustrated in Fig. 2, we model this aspect by assuming that the samples $Y_i$ are not directly available at the decoder, but are instead measured by the latter with a quality that can be controlled by an "action" variable $A_i$. Selection of the action sequence $A^n$ by the decoder has to satisfy given cost constraints. This model is inspired by the "vending machine" setting proposed in [6], where a problem similar to Fig. 2 was studied where the decoder is only interested in estimating $X^n$ (and not a function $f^n(X^n, Y^n, Z^n)$). We refer to the setting in Fig. 2 as *lossy computing with observation costs*.

2) *Robust computing*: In sensor networks and cloud computing, reliability of all the computing devices (e.g., sensors or servers) cannot be guaranteed all the time. Therefore, it is appropriate to design the system so as to be robust to system failures. As shown in Fig. 3, we model this aspect by assuming that the decoder, unbeknownst to the encoder, may not be able to acquire information about the sequence $Y^n$. This setting is equivalent to assuming the presence of two decoders, one with the capability to acquire information about $Y^n$ (Decoder 2) and one without this capability (Decoder 1). This model

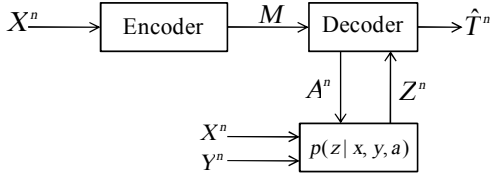

Fig. 2. Lossy compuitng with observation costs where decoder wishes to compute a function $T^n = \mathrm{f}^n(X^n, Y^n, Z^n)$ with $T_i = \mathrm{f}(X_i, Y_i, Z_i)$ for $i \in [1, n]$.

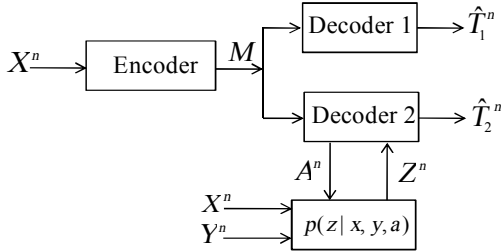

Fig. 3. Lossy compuitng with observation costs where decoder 1 wishes to compute a function $T_1^n = \mathrm{f}_1^n(X^n, Y^n, Z^n)$ with $T_{1i} = \mathrm{f}_1(X_i, Y_i, Z_i)$ for $i \in [1, n]$ and decoder 2 wishes to compute a function $T_2^n = \mathrm{f}_2^n(X^n, Y^n, Z^n)$ with $T_{2i} = \mathrm{f}_2(X_i, Y_i, Z_i)$ for $i \in [1, n]$

is inspired by the so called Heegard-Berger-Kaspi problem [7][8], where the two decoders of Fig. 3 are interested in estimating $X^n$ and the measurement of $Y^n$ at decoder 2 is perfect (i.e., $Z^n = Y^n$). We refer to the setting in Fig. 3 as *robust lossy computing with observation costs*.

Our main contributions are as follows. For the robust problem of lossy coding with observation costs of Fig. 3, and thus as a special case for the problem in Fig. 2, we derive the rate-distortion-cost function (to be defined) and obtain conclusive results on this function in special cases of interest in Sec. III. Moreover, in Sec. III-A, we present an example with binary sources to discuss the impact of observations costs. Finally, in Sec. IV, we present extensions of the considered model to the case where the side information is obtained in a causal way at the decoder.

*Notation*: Upper case, lower case and calligraphic letters denote discrete random variables, specific values of random variables and their alphabets, respectively. For integers $a \leq b$, we define $[a, b]$ to be set of all integers between $a$ and $b$ (i.e., $a, a+1, ..., b$). Finally, $\delta(x)$ represents the Kronecker delta function.

## II. SYSTEM MODEL

In this section, the system model for the problems illustrated in Fig. 2 and Fig. 3 are formalized in Sec. II-A and Sec. II-B, respectively.

### A. Lossy Computing with Observation Costs

The problem of lossy computing with observation costs, illustrated in Fig. 2, is defined by the probability mass functions (pmfs) $p_{XY}(x, y)$ and $p_{Z|XYA}(z|x, y, a)$ and corresponding alphabets as follows. The source sequences $X^n$ and $Y^n$, with $X_i \in \mathcal{X}$ and $Y_i \in \mathcal{Y}$ for $i \in [1, n]$, are such that the pairs $(X_i, Y_i)$ are independent identically distributed with joint probability mass function (pmf) $p_{XY}(x, y)$. The encoder measures sequence $X^n$ and encode it in a message $M$ of $nR$ bits, which is delivered to the decoder. The decoder wishes to estimate a sequence $T^n = \mathrm{f}^n(X^n, Y^n, Z^n)$, with $T_i = \mathrm{f}(X_i, Y_i, Z_i)$ for $i \in [1, n]$, for a given function f: $\mathcal{X} \times \mathcal{Y} \times \mathcal{Z} \to \mathcal{T}$. To this end, the decoder receives message $M$ and, based on this, selects an action sequence $A^n$, where $A_i \in \mathcal{A}$ for $i \in [1, n]$. The action sequence affects the quality of the measurement $Z^n$ of sequences $X^n$ and $Y^n$ obtained at the decoder. Specifically, given $A^n$, $X^n$ and $Y^n$, the sequence $Z^n$ is distributed as

$$p(z^n|x^n, y^n, a^n) = \prod_{i=1}^{n} p_{Z|XYA}(z_i|x_i, y_i, a_i). \qquad (1)$$

The cost of an action sequence $a^n$ is defined by a cost function $\Lambda: \mathcal{A} \to [0, \infty)$ as $\Lambda^n(a^n) = 1/n \sum_{i=1}^n a_i$. The estimated sequence $\hat{T}^n$ with $\hat{T}_i \in \hat{\mathcal{T}}$ for $i \in [1, n]$ is then obtained as a function of $M$ and $Z^n$. Let $d: \mathcal{T} \times \hat{\mathcal{T}} \to [0, \infty)$ be a distortion measure. The distortion between the desired sequence $t^n$ and the reconstruction $\hat{t}^n$ is defined as $d^n(t^n, \hat{t}^n) = 1/n \sum_{i=1}^n d(t_i, \hat{t}_i)$. A formal description of the operations at encoder and decoder is presented below.

*Definition 1:* An $(n, R, D, \Gamma)$ code for lossy computing with observation costs (Fig. 2) consists of a source encoder

$$\mathrm{g}: \mathcal{X}^n \to [1, 2^{nR}], \qquad (2)$$

which maps the sequence $X^n$ into a message $M$; an "action" function

$$\ell: [1, 2^{nR}] \to \mathcal{A}^n, \qquad (3)$$

which maps the message $M$ into an action sequence $A^n$; and a decoding function

$$\mathrm{h}: [1, 2^{nR}] \times \mathcal{Z}^n \to \hat{\mathcal{T}}^n, \qquad (4)$$

which maps the message $M$ and the measured sequence $Z^n$ into the estimated sequence $\hat{T}^n$; such that the action cost constraint $\Gamma$ and distortion constraint $D$ are satisfied, i.e.,

$$\frac{1}{n}\sum_{i=1}^{n} \mathrm{E}\left[\Lambda(A_i)\right] \leq \Gamma \qquad (5)$$

and 
$$\frac{1}{n}\sum_{i=1}^{n} \mathrm{E}\left[d(T_i, \hat{T}_i)\right] \leq D, \qquad (6)$$

respectively.

*Definition 2:* Given a distortion-cost pair $(D, \Gamma)$, a rate $R$ is said to be achievable if, for any $\epsilon > 0$, and sufficiently large $n$, there exists a $(n, R, D + \epsilon, \Gamma + \epsilon)$ code.

*Definition 3:* The *computational rate-distortion-cost function* $R(D, \Gamma)$ is defined as $R(D, \Gamma) = \inf\{R: \text{the triple } (R, D, \Gamma) \text{ is achievable}\}$.

*Remark 1:* The system at hand reduces to several settings studied in the literature. If $p(z|a, x, y) = p(z|x)$, and the decoder wishes to estimate $X^n$ (i.e., $\mathrm{f}(x, y, z) = x$), the system of Fig. 2 becomes the standard Wyner-Ziv problem [9]. More in general, if $p(z|a,x,y) = p(z|x)$ and $f(x,y,z) = f(x,z)$, the problem becomes the one studied by Yamamoto in [3]. Finally, if we set $X_i = Y_i$ and $f(x,y,z) = x$, the problem reduces to the setting of lossy source coding with a vending machine studied in [6].

### B. Robust Lossy Computing with Observation Costs

The setting of robust lossy computing with observation costs, illustrated in Fig. 3, generalizes the setting in Fig. 2. In fact, here, decoder 2 is defined as the decoder in Fig. 2, and is thus interested in estimating a function $T_2^n = f_2^n(X^n, Y^n, Z^n)$ with $T_{2i} = f_2(X_i, Y_i, Z_i)$ for $i \in [1,n]$. However, message $M$ is also received by decoder 1, that does not have access to any further measurement and wishes to calculate a sequence $T_1^n = f_1^n(X^n, Y^n, Z^n)$ where $T_{1i} = f_1(X_i, Y_i, Z_i)$ for $i \in [1,n]$. The problem for the encoder is to cater to both decoders, thus obtaining a performance that is robust to uncertainties about the availability of the side information $Z^n$. The code definition follows similarly to the problem of lossy computing with observations costs in Definition 1, where we fix two distortion functions $d_1: \mathcal{T} \times \hat{\mathcal{T}}_1 \to [0,\infty)$ and $d_2: \mathcal{T} \times \hat{\mathcal{T}}_1 \to [0,\infty)$ for decoder 1 and decoder 2, respectively.

*Definition 4:* An $(n, R, D_1, D_2, \Gamma)$ code for robust lossy computing with observation costs (Fig. 3) consists of a source encoder (2); an action function (3); a decoding function for decoder 1
$$h_1: [1, 2^{nR}] \to \hat{\mathcal{T}}_1^n, \quad (7)$$
which maps the message $M$ into the estimate $\hat{T}_1^n$; a decoding function for decoder 2
$$h_2: [1, 2^{nR}] \times \mathcal{Z}^n \to \hat{\mathcal{T}}_2^n, \quad (8)$$
which maps the message $M$ and the sequence $Z^n$ into the estimate $\hat{T}_2^n$; such that the action cost constraint (5) is satisfied and the distortion constraints
$$\frac{1}{n}\sum_{i=1}^n \mathrm{E}\left[d_j(T_{ji}, \hat{T}_{ji})\right] \leq D_j \quad (9)$$
hold for $j = 1, 2$.

Achievability of code for a triple $(D_1, D_2, \Gamma)$ and the *computational rate-distortion-cost function* $R(D_1, D_2, \Gamma)$ are defined as above.

*Remark 2:* If $p(z|a,x,y) = p(z|x)$ and both decoders wish to estimate $X$ (i.e., $f_1(x,y,z) = f_2(x,y,z) = x$), the system of Fig. 3 reduces to the so called Heegard-Berger-Kaspi problem [7] and [8]. The binary-alphabet version of this problem was studied in [10], as further discussed below.

## III. ROBUST LOSSY COMPUTING WITH OBSERVATION COSTS

In this section, we derive the computational rate distortion-cost function defined in Sec. II-B for robust lossy computing with observation costs of Fig. 3. Corresponding results for the model of Fig. 2 are obtained by setting the distortion constraint $D_1$ to be larger than or equal to $d_{1\max} = \min_{x \in \hat{\mathcal{T}}_1} \mathrm{E}[d_1(T_1, x)]$, or equivalently by setting $f_1(x)$ equal to a constant $x \in \hat{\mathcal{T}}_1$. We drop the subscripts from the pmfs for simplicity of notation (e.g., $p_{XY}(x,y)$ is defined as $p(x,y)$).

*Proposition 1:* The computational rate-distortion-cost function for the system with two decoders (Fig. 3) is givens by
$$R(D_1, D_2, \Gamma) = \min_{\substack{p(a,u,\hat{t}_1|x),\, \hat{t}_2(u,z):\\ \mathrm{E}[\Lambda(A)] \leq \Gamma \text{ and } \mathrm{E}[d_j(T_j,\hat{T}_j)] \leq D_j, j=1,2}} [I(X;A)$$
$$+ I(X; \hat{T}_1|A) + I(X; U|Z, A, \hat{T}_1), \quad (10)$$
where the mutual informations are calculated with respect to the distribution
$$p(x,y,z,u,a,\hat{t}_1,\hat{t}_2) = p(x,y)p(z|a,x,y) \quad (11)$$
$$\cdot p(a,u,\hat{t}_1|x)\delta(\hat{t}_2 - \hat{t}_2(u,z))$$
and $\hat{t}_2(u,z)$ is a (deterministic) function. Finally, in (10), $U$ is an auxiliary random variable whose alphabet has cardinality $|\mathcal{U}| \leq |\mathcal{X}||\mathcal{A}| + 3$ without loss of optimality.

The achievable rate (10) can be easily interpreted. The encoder first maps the input sequence $X^n$ into an action sequence $A^n$ so that the two sequences are jointly typical, which requires $I(X;A)$ bit/source sample. Then, it maps $X^n$ into the estimate $\hat{T}_1^n$ for decoder 1 using a conditional codebook with rate $I(X; \hat{T}_1|A)$. Finally, it maps $X^n$ into another sequence $U^n$ using the fact that decoder 2 has the action sequence $A^n$, the estimate $\hat{T}_1^n$ and the measurement $Z^n$. Using conditional codebooks (with respect to $\hat{T}_1^n$ and $A^n$) and from the Wyner-Ziv theorem, this requires $I(X; U|A, Z, \hat{T}_1)$ bit/source sample (see, e.g., [2]). Decoder 2 then estimates $\hat{T}_2^n$ sample by sample by using a function $\hat{T}_{2i} = \hat{t}_2(U_i, Z_i)$.

*Remark 3:* By setting $D_1 \geq d_{1\max}$, or $T_1$ equal to a constant, corresponding results can be derived for the setting of Fig. 2, where $D = D_2$. Specifically, with $T_1$ (and thus $\hat{T}_1$) constant, the result in Proposition 1 recovers Theorem 1 in [3] for the special case in which $p(z|a,x,y) = p(z|x)$ and the decoder wishes to compute a function $f(x,z)$. Moreover, the result in Proposition 1 recovers Theorem 1 of [6] by setting $f(x,y,z) = x$.

*Remark 4:* For the scenario in Fig. 3, the result in Proposition 1 recovers the rate-distortion function for the Heegard-Berger-Kaspi problem [7][8] if $p(z|a,x,y) = p(z|x)$ and both decoders wish to estimate $X$ (i.e., $f_1(x,y,z) = f_2(x,y,z) = x$).

*Remark 5:* In the results of this section, one can set $A$ to be a deterministic function $a(U)$ of $U$ without loss of optimality. As in [6], this can be seen by redefining $U$ as $(U, A)$ and noticing that, with this new definition, mutual informations and the joint distribution in the results above remain the same.

### A. A Binary Example

We now focus on the specific example for the setting in Fig. 2, where all variables are binary and $(X, Y)$ is a doubly symmetric binary source (DSBS) characterized by probability $p$, $0 \leq p \leq 1/2$, so that $\Pr[X \neq Y] = p$. Moreover, the decoder wishes to compute a function $T = f(X, Y, Z) = X \otimes Y$, where $\otimes$ denotes the binary multiplication of $X$ and $Y$.

The measurement of $Y_i$ takes place over a system $p(z|a, x, y)$ defined as

$$Z_i = \begin{cases} Y_i & \text{if } A_i = 1 \\ 1 & \text{if } A_i = 0 \end{cases}, \quad (12)$$

so that when the decoder sets $A_i = 1$, sample $Y_i$ is measured, while otherwise it is not. In selecting the action $A_i$, the decoder must satisfy the cost constraint defined as $\Lambda(A_i) = 1$ if $A_i = 1$ and $\Lambda(A_i) = 0$ if $A_i = 0$, so that, given (12), a cost $\Gamma$ implies that the decoder can observe $Y^n$ only for at most $n\Gamma$ symbols. Finally, we take the Hamming distortion as the distortion metric, i.e., $d(t_j, \hat{t}_j) = 0$ if $t_j = \hat{t}_j$ and $d(t_j, \hat{t}_j) = 1$.

We start by observing that, if the cost is $\Gamma = 1$, the decoder can set $A_i = 1$ at all times $i \in [1, n]$ and thus measure the entire sequence $Z^n = Y^n$, so that the problem of obtaining the rate-distortion function reduces to the one studied, and solved, in [3]. In [3], it is also shown that, among binary functions, focusing on the Hamming distortion metric and DSBS, calculation of the binary product is the most challenging. In fact, calculation of other binary functions, such as the binary sum, reduces to either trivial problems or the standard Wyner-Ziv problem. This is why we focus here on the binary product.

Let us make some further general observations on the rate-distortion-cost function for this example. First, if $D \geq (1-p)/2$, then $R(D, \Gamma) = 0$ for all values of $\Gamma \geq 0$. The reason is that, if the decoder always reconstructs $\hat{T} = 0$, which requires zero rate, the Hamming distortion equals $\Pr[X \otimes Y = 1] = (1-p)/2$. Moreover, if $(1-p)/2 \geq D \geq p/2$, then $R(D, \Gamma) = 0$ for a $\Gamma \leq 1$ sufficiently large. In fact, assume that $\Gamma = 1$ so that $Z = Y$. The decoder can now reconstruct $\hat{T} = 0$ if $Y = 0$ with no distortion, and $\hat{T} = 1$ if $Y = 1$, thus incurring a distortion equal to $p$. The overall average distortion is $p/2$. Therefore, any distortion $D$ larger that $p/2$ can also be achieved at a cost generally smaller than $\Gamma = 1$ by, for instance, setting $A_i = 0$ with probability $\frac{D-p/2}{1-p/2}$ and by using the same arguments above. In fact, the distortion with this choice is easily seen to be upper bounded by $\frac{D-p/2}{1-p/2} + (1 - \frac{D-p/2}{1-p/2}) \cdot \frac{p}{2} = D$.

For a general cost $\Gamma \leq 1$ and distortion $D$, computational rate-distortion-cost can be found by solving (10) (with $T_1$ and $\hat{T}_1$ equal to a constant and $D_2 = D$). The optimization in (10) is in general not easy. However, in [3], it is shown that, if $p(z|x, y, a) = p(z|x)$, and thus in our model if $\Gamma = 1$, a binary $U$, i.e., $|\mathcal{U}| = 2$, and a decoding function $\hat{t}(u, z) = u \otimes z$ are optimal. Based on this result, here we optimize (10) under the constraints that $|\mathcal{U}| = 2$ and $\hat{t}(u, z) = u \otimes z$ without claiming that it maximizes the computational rate-distortion-cost (10).

Fig. 4 shows the calculated computational rate-distortion-cost for $p = 0.2$ and $D = 0.05, 0.15$ versus the cost $\Gamma$. It can be seen, as discussed above that for $D = 0.15$, which satisfies $(1-p)/2 \geq D \geq p/2$, if the cost is large enough, the rate is zero, whereas for $D = 0.05 < p/2$, this is not the case.

A final remark is in place. In the model at hand, the actions $A_i$ are selected as a function of the message $M$. In (10), this amounts to choosing a distribution $p(a|x)$ that depends on $x$. A simpler approach, referred to as "greedy" in [6] is to choose $A_i$ independently of $M$. This way, the decoder does not wait to hear from the encoder before choosing how to act on the measurement system that records $Z_i$. The corresponding rate can be obtained from (10) by setting $A$ to be independent of $X$. Such rate, calculated assuming again $|\mathcal{U}| = 2$ and $\hat{t}(u, z) = u \otimes z$, is shown for comparison in Fig. 4. It can be seen that, while for $\Gamma = 1$ the greedy approach is clearly optimal (as one can set $A = 1$), for all other costs it is generally suboptimal. This shows the importance of coordinating the measurement system at the decoder with the encoder.

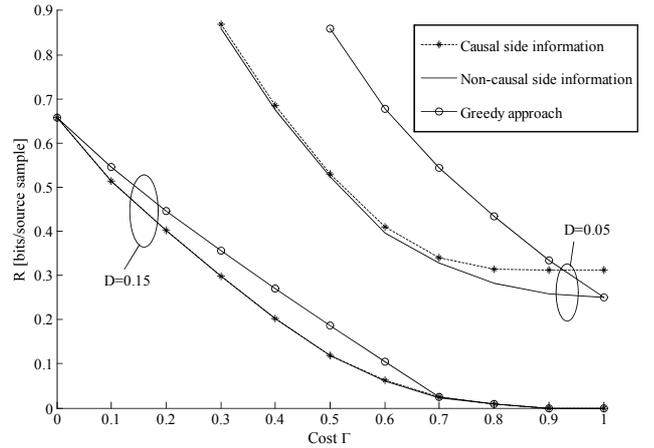

Fig. 4. Computational rate-distortion-cost for the binary example of Sec. III-A ($p = 0.2$).

## IV. CAUSAL SIDE INFORMATION

In this section, decoder 2 needs to make a decision about $\hat{T}_{2i}$ after having observed only the first $i$ samples of $Z^n$, i.e., $Z^i$, instead of the full sequence $Z^n$ as assumed above. In this model, a code is defined as in Definition 4 with the only difference that the decoding function (8) is given by

$$\text{h}_{2i}: [1, 2^{nR}] \times \mathcal{Z}^i \to \hat{\mathcal{T}}_2, \quad (13)$$

for $i \in [1, n]$, which maps the message $M$ and the sequence $Z^i$ to the reconstructed symbol $\hat{T}_{2i} = \text{h}_{2i}(M, Z^i)$.

*Proposition 2:* The computational rate-distortion-cost function for robust lossy computing with causal observation at decoder 2 is

$$R(D_1, D_2, \Gamma) = \min_{\substack{p(a,u,\hat{t}_1|x), \hat{t}_2(u,z): \\ \text{E}[\Lambda(A)] \leq \Gamma \text{ and } \text{E}[d_j(T_j, \hat{T}_j)] \leq D_j, j=1,2}} [I(X; A)$$
$$+ I(X; \hat{T}_1|A) + I(X; U|A, \hat{T}_1)], \quad (14)$$

where the joint distribution factorizes as

$$p(x, y, z, u, a, \hat{t}_1, \hat{t}_2) = p(x, y)p(z|a, x, y)p(a, u, \hat{t}_1|x)$$
$$\cdot \delta(\hat{t}_2 - \hat{t}_2(u, z)), \quad (15)$$

$\hat{t}_2(u, z)$ is a (deterministic) functions, and $U$ is an auxiliary random variable with cardinality $|\mathcal{U}| \leq |\mathcal{X}||\mathcal{A}| + 3$.

It is noted that the main difference in the achievability of Proposition 1 and Proposition 2 is that, in the latter, the encoder cannot employ Wyner-Ziv encoding when conveying the sequence $U^n$ to decoder 2, since the sequence $Z^n$ is available only causally to decoder 2. This modifies the third term in (14) as compared to (10) from $I(X; U|A, Z, \hat{T}_1)$ to $I(X; U|A, \hat{T}_1)$ and thus generally increases the rate-distortion-cost function.

*Remark 6:* In case decoder 1 is not present (or equivalently $D_1$ is sufficiently large), the scenario at hand was studied in [6], where the rate-distortion-cost function was derived for the case where $f_2(x, y, z) = x$, that is, the decoder wishes to reconstruct $X$. Proposition 2 thus generalizes the aforementioned result in [6].

*Remark 7:* Similar to Remark 5, in (14) one can set $A$ as a function of $U$ and $\hat{T}_1$ as a function of $A$ without loss of optimality.

*Example 1:* We continue the example in Sec. III-A and evaluate the rate-distortion-cost function with causal side information. This can be obtained from Proposition 2. As in Sec. III-A, to simplify the evaluation of (14), we assume, without claiming optimality, that $U$ is binary and $\hat{t}(u, z) = u \otimes z$. The corresponding rate-distortion-cost function is shown in Fig. 4. It can be seen that given a fixed value of $p$, the smaller the value of $D$, the larger is the gap between the computational-rate-distortion-cost function for causal side information with respect to that with non-causal side information. Moreover, the difference between the rates becomes negligible by increasing the value of $D$. This can be explained since for larger values of $D$, the advantage of using non-causal side information over causal side information decreases. Besides, for small values of $D$ and given a fixed value of $p$, the gap between the rate corresponding to the causal and non-causal side information becomes larger by increasing the value of cost, $\Gamma$.

## V. Concluding Remarks

Minimizing the transmission rate required for the computation of given functions of data available at distinct nodes is a problem of interest in many applications including sensor networks and cloud computing. In this work, we have studied a standard baseline scenario for this problem by addressing two key issues: (*i*) Acquiring the data necessary for computation may be costly and thus should be done efficiently; (*ii*) Failures may occur in remote nodes that cannot be predicted by other devices. We have included these two aspects in an information-theoretic formulation that we referred to as robust lossy computing with observations costs (see Fig. 3). Conclusive results are reported, along with numerical examples.